# Ultrafast enhancement of ferromagnetic spin exchange induced by ligand-to-metal charge transfer

A. Ron[1,2], S. Chaudhary[1,2], G. Zhang[3], H. Ning[1,2], E. Zoghlin[4], S. D. Wilson[4], R. D. Averitt[3], G. Refael[1,2] & D. Hsieh[1,2]

[1]Department of Physics, California Institute of Technology, Pasadena, California 91125, USA

[2]Institute for Quantum Information and Matter, California Institute of Technology, Pasadena, California 91125, USA

[3]Department of Physics, University of California, San Diego, La Jolla, California 92093, USA

[4]Materials Department, University of California, Santa Barbara, California 93106, USA

**Directly modifying spin exchange energies in a magnetic material with light can enable ultrafast control of its magnetic states. Current approaches rely on tuning charge hopping amplitudes that mediate exchange by optically exciting either virtual[1–8] or real[9–12] charge-transfer transitions (CT) between magnetic sites. Here we show that when exchange is mediated by a non-magnetic ligand, it can be substantially enhanced by optically exciting a real CT transition from the ligand to magnetic site, introducing lower order virtual hopping contributions. We demonstrate sub-picosecond enhancement in a superexchange dominated ferromagnet $CrSiTe_3$ through this mechanism using phase-resolved coherent phonon spectroscopy. This technique can also be applied in the paramagnetic phase to disentangle light induced exchange modification from other ultrafast effects that alter the magnetization. This protocol can potentially be broadly applied to engineer thermally inaccessible spin Hamiltonians in superexchange dominated magnets.**

In superexchange dominated magnetic insulators, as is realized in many transition metal oxide and chalcogenide based materials, the superexchange energy $J_{ex}$ between spins on neighboring metal sites depends on the number of virtual hopping processes that mediate the interaction. A simple illustrative model consists of two metal ions, each with a singly occupied *d*-orbital, interacting via a ligand ion with fully occupied $p_x$ and $p_y$ orbitals in a 90° geometry (Fig. 1a). To leading order in perturbation theory $J_{ex} \propto -\left(\frac{t}{\Delta_{CT}}\right)^4 J_H$, where $t$ is the metal-to-ligand hopping energy, $\Delta_{CT}$ is the charge transfer energy between the ligand *p*- and metal *d*-orbital, and $J_H$ is the Hund's coupling between the *p*-orbitals on the ligand site (see Supplementary Section S1 for details). This expression reflects the fact that the superexchange is ferromagnetic and is only sensitive to fourth-order virtual hopping processes. In contrast, for a CT excited state where an unpaired spin on a metal site moves to the ligand site (Fig. 1a), the leading order contribution is sensitive to second-order virtual hopping processes owing to the Hund's coupling on the ligand site and is given by $J_{ex} \propto -\left(\frac{t^2}{\Delta_{CT}^2 - J_H^2}\right) J_H$. Since $\Delta_{CT}$ can be several times larger than both $t$ and $J_H$ in these materials, one can in principle transiently enhance $J_{ex}$ by exciting a CT transition with an ultrafast laser pulse.

Currently, there are two main experimental approaches to measuring ultrafast laser induced exchange ($\delta J_{ex}$) modification. One is to detect $\delta J_{ex}$ induced coherent spin precession using THz emission[2] or magneto-optical Kerr spectroscopy[9]. Another is to resolve the transient renormalization of spin wave energies or exchange splitting energies using techniques such as femtosecond stimulated Raman scattering[3] or photoemission spectroscopy[13,14] respectively. We propose an alternative approach based on measuring $\delta J_{ex}$ induced displacive excitation of coherent phonons, dubbed spin-DECP. In conventional DECP[15], an ultrafast optical pulse instantaneously modifies the charge distribution in a material, which shifts the atomic coordinates of the potential energy minimum thereby producing a restoring Coulombic force $\vec{F}_C$ that initiates oscillatory lattice motion. In a superexchange system, the equilibrium metal-ligand-metal bonding angle $\theta$ is determined by a balance between the total elastic and total magnetic energies. The latter typically has the form $\sum_{ij} J_{ex}(\theta) \langle \vec{S}_i \cdot \vec{S}_j \rangle$, where $\langle \vec{S}_i \cdot \vec{S}_j \rangle$ is the spin correlator between metal sites $i$ and $j$. An instantaneous change in magnetic energy due to $\delta J_{ex}$ shifts the potential energy minimum away from the equilibrium $\theta$ value, producing a restoring exchange force $\vec{F}_{ex}$ that initiates oscillatory motion (Fig. 1b). The advantage of using spin-DECP to detect $\delta J_{ex}$ over the aforementioned techniques is that it relies only on the existence of short-range ($\langle \vec{S}_i \cdot \vec{S}_j \rangle \neq 0$) and not long-range magnetic correlations ($\langle \vec{S}_i \rangle \neq 0$). This helps to disentangle $\delta J_{ex}$ from the many laser-induced processes that are known to affect magnetic order[16], and it can be applied to both antiferromagnetic (AFM) and ferromagnetic (FM) materials.

An ideal material for demonstrating ultrafast ligand-to-metal CT induced enhancement of $J_{ex}$ using spin-DECP is $CrSiTe_3$, a prototypical superexchange FM insulator with a layered honeycomb structure. In this material the magnetic ($S = 3/2$) degrees of freedom originate from half-filled Cr (metal) $3d$ $t_{2g}$ orbitals, which interact predominantly with its nearest neighbors[17] within the honeycomb layer via FM superexchange mediated by the Te (ligand) 5$p$-orbitals. Direct AFM exchange between Cr $t_{2g}$ orbitals also exists but is expected to be weak due to the large Cr-Cr distance[18]. Owing to its quasi-two-dimensional structure, short-range intra-layer FM correlations in CrSiTe$_3$ persist up to ~ 110 K[17,19], far above the Curie temperature $T_c$ = 33 K. Based on optical absorption[18] and density functional theory (DFT) calculations[20,21], optical excitation across the direct bandgap of CrSiTe$_3$ primarily involves CT from the Te $p$- to Cr $e_g$-orbital (Fig. 2a), which, within our single $d$-orbital model, potentially leads to a ~5-fold enhancement of $J_{ex}$ using reported values of $\Delta_{CT}$ ~ 1 eV and $t$ ~ 0.5 eV for CrSiTe$_3$[21] and a typical value of $J_H$ ~ 0.5 eV[22]. The multi $d$-orbital nature of the magnetism in CrSiTe$_3$ does not alter the conclusions drawn from our single $d$-orbital model (Fig. 1a) because the energies of the additional second-order virtual hopping processes introduced do not depend on the relative spin on the Cr sites (see Supplementary Section S1). We note that CT excitation is also expected to weaken the AFM direct exchange, thus further enhancing the net FM exchange. This is because Coulomb repulsion from an additional electron in the $e_g$ orbital suppresses virtual hopping between Cr $t_{2g}$ orbitals, and because virtual hopping from a half-filled to empty $e_g$ orbital gives rise to a FM exchange according to the Goodenough-Kanamori rules[23–25].

To distinguish spin-DECP from conventional DECP effects, which is generally challenging because both contribute to exciting a coherent phonon[10], we target modes whose oscillation phase is sensitive to the excitation mechanism. An example is the $A_g^2$ optical phonon mode in CrSiTe$_3$, which involves the periodic expansion and contraction of the Te octahedra that surround each Cr ion (Fig. 2a). An optical excitation that transfers an electron from the Te$^{2-}$ to Cr$^{3+}$ ions launches this mode through DECP, primarily by reducing the electrostatic attraction between them, producing a repulsive $\vec{F}_C$ that causes an initial expansion of the Te octahedra. Importantly, the same CT excitation also increases $J_{ex}$, which launches the mode through spin-DECP by pushing $\theta$ towards 90° where the FM superexchange is strongest according to the Goodenough-Kanamori rules[23–25]. Since $\theta$ < 90° in equilibrium[26], this produces an attractive $\vec{F}_{ex}$ that causes an initial contraction of the Te octahedra. Thus DECP and spin-DECP drive the $A_g^2$ mode with opposite initial phase.

Phase-resolved coherent phonon spectroscopy of a bulk CrSiTe$_3$ single crystal was performed using ultrafast pump-probe optical reflectivity measurements (Fig. 2a). A typical reflectivity transient acquired using a pump photon energy resonant with $\Delta_{CT}$ is shown in Figure 2b, revealing an abrupt drop at delay time $t = 0$ followed by an exponential recovery on a $\tau \sim 2$ ps timescale, tracking the generation and relaxation of CT excitations respectively. The periodic modulations atop the recovery arise from a coherent optical phonon with frequency $f \sim 3.8$ THz, which can be assigned to the $A_g^2$ mode based on Raman spectroscopy and DFT calculations[18,27,28]. A comparison of the oscillatory component ($\Delta R/R_{osc}$), isolated by subtracting the exponential background from the data, acquired above and below $T_c$ reveal a $\pi$ phase difference (Fig. 2b). This suggests that $A_g^2$ mode excitation evolves from conventional DECP to spin-DECP dominated upon cooling.

To verify a spin-DECP mechanism at low temperature in CrSiTe$_3$, we note that pronounced growth of 2D FM correlations was detected below a temperature scale $T_{2D} \approx 110$ K by both neutron scattering[17] and optical second harmonic generation[19]. Therefore one should expect spin-DECP effects to be present well above $T_c$. A complete temperature dependence of $\Delta R/R_{osc}$ and the optical second harmonic susceptibility of our sample are displayed in Figures 3a & b. The phonon oscillation phase $\phi$ extracted from the $\Delta R/R_{osc}$ data (see Supplementary Section S2 for extraction procedure) undergoes a sharp change from 0 to $\pi$ below $T \sim 90$ K (Fig. 3c), well inside the 2D correlated paramagnetic regime. To understand this behavior, we re-analyze the data with $\phi$ absorbed into the sign of the oscillation amplitude $A_{ph}$. This procedure is valid because $\phi$ must be either 0 or $\pi$, corresponding to positive and negative signs for $A_{ph}$, and helps visualize shifts in the relative weight of the DECP and spin-DECP contributions. As shown in Figure 3c, $A_{ph}$ is positive ($\phi = 0$) and temperature independent for $T > T_{2D}$, consistent with the $A_g^2$ mode being driven exclusively by DECP. Below $T_{2D}$ where $\langle \vec{S}_i \cdot \vec{S}_j \rangle$ begins to grow, a spin-DECP contribution with negative amplitude ($\phi = \pi$) starts to compete with DECP, causing $A_{ph}$ to decrease and eventually change sign at $T \sim 90$ K where $\vec{F}_{ex}$ becomes equal and opposite to $\vec{F}_C$ (see Supplementary Section S1 for numerical estimates of $\vec{F}_{ex}$ and $\vec{F}_C$). This crossover temperature has no measureable pump fluence dependence as expected since both the total $\vec{F}_{ex}$ and $\vec{F}_C$ should scale with the CT excitation density. As a control, we also measured coherent oscillations of an $E_g$ phonon under identical experimental conditions. In contrast to the $A_g^2$ mode, the $E_g$ mode can only be excited via impulsive stimulated Raman scattering and not conventional or spin-DECP due to symmetry considerations. We verified that both the amplitude and phase of the $E_g$ mode are indeed temperature independent (see Supplementary Section S3).

Below 90 K, $A_{\text{ph}}$ becomes increasingly negative as $\langle \vec{S}_i \cdot \vec{S}_j \rangle$ continues to rise (Fig. 3a). However $A_{\text{ph}}$ reaches a minimum around $T_c$ and then turns back towards zero upon further cooling despite $\langle \vec{S}_i \cdot \vec{S}_j \rangle$ continuing to increase. This trend can be explained by a weakening of the spin-DECP contribution due to exchange striction below $T_c$. Previous x-ray diffraction studies[26] on CrSiTe$_3$ showed that below $T_c$ there is a continual change in lattice parameters that brings $\theta$ closer to 90°. This means that the separation in $\theta$ between the potential energy minima of the equilibrium and CT states become smaller (Fig. 1b), progressively reducing $\vec{F}_{ex}$. Exchange striction must dominate the competing effect of increasing $\langle \vec{S}_i \cdot \vec{S}_j \rangle$, leading to a net decreasing spin-DECP to DECP ratio upon cooling below $T_c$.

To uniquely demonstrate that the observed ultrafast enhancement of $J_{ex}$ is induced by a ligand-to-metal CT transition, we repeated our coherent phonon spectroscopy measurements with a pump photon energy of 0.14 eV, far below the indirect (~0.4 eV) bandgap of CrSiTe$_3$[18], to suppress CT transitions between the Te and Cr ions. We note that the absorption edge of CrSiTe$_3$ is characterized by a long Urbach tail[29] that extends well below the indirect gap[18], allowing for finite absorption even at 0.14 eV. Given that the highest energy phonon mode in CrSiTe$_3$ lies at 0.06 eV[18], absorption at such low photon energies is likely to be dominated by transitions into mid-gap defect states[30] rather than phonon assisted CT processes. In fact, DFT calculations have shown that it is energetically favorable to form Cr/Si anti-site defects in CrSiTe$_3$[30] and scanning tunneling microscopy measurements on the closely related compound CrGeTe$_3$ have shown that this type of defect produces a finite density of in-gap states[31]. An optical CT excitation from the Te $p$- to Si defect orbitals should produce a Coulombic force between the Te and Cr atoms in the same direction as in the Te-Cr CT state due to increased charge negativity of the Te site. However, because the spin of the photo-hole on the Te site is independent of the spin on the Cr sites, second-order virtual hopping processes in the Te-Si CT state do not contribute to $J_{ex}$. In line with this scenario, Figure 4 shows that a 0.14 eV pump is indeed able to excite the $A_g^2$ mode well above $T_c$ via conventional DECP but that no phase change occurs down to 7 K (see Supplementary Section S4 for additional data). We note that while a renormalization of $J_{ex}$ imparted by a photo-assisted virtual hopping (i.e. Floquet engineering) mechanism can in principle also drive spin-DECP, such effects are negligible (< 1 %) for the pump electric fields (~ 0.1 V/Å) used in our experiments (see Supplementary Section S5 for detailed calculation).

The ability to significantly enhance nearest-neighbor superexchange via ligand-to-metal CT excitation as we have demonstrated can potentially be harnessed to transiently manipulate magnetic materials in new ways. Examples include optically inducing FM order above $T_c$, switching between FM and AFM ordering patterns by altering the ratio of nearest-neighbor to further neighbor interactions, which are known to compete for example in transition-metal trichalcogenide materials such as CrSiTe$_3$[32,33], or inducing order in geometrically frustrated magnets by controlling the light polarization to generate anisotropic changes in $J_{ex}$ that relieve frustration. Optically induced FM ordering above $T_c$ was not achievable in our experiments because the stabilization of long-range FM ordering in CrSiTe$_3$ depends predominantly on inter-layer exchange[17,19], which is not enhanced by our pump. Moreover, owing to the intra- and inter-layer exchange time scales in CrSiTe$_3$ ($h/J_{ex}$ ~ 4 ps) being longer than the electron-hole recombination time ($\tau$ < 2 ps according to Fig. 2b), there is likely insufficient time for global spin rearrangement. Realizing the aforementioned vision will therefore involve targeting exchange pathways in materials that most strongly affect magnetic order, engineering both materials and excitation protocols to extend $\tau J_{ex}/h$, and developing *ab initio* approaches to calculate $J_{ex}$ in the CT state of complex solids to refine the molecular toy model estimates considered in this work.

**Acknowledgements**

This work was supported by ARO MURI Grant No. W911NF-16-1-0361. D.H. and G.R also acknowledge support from the David and Lucile Packard Foundation. D.H. acknowledges support for instrumentation from the Institute for Quantum Information and Matter, an NSF Physics Frontiers Center (PHY-1733907). A.R. acknowledges support from the Caltech Prize Fellowship. The MRL Shared Experimental Facilities are supported by the MRSEC Program of the NSF under Award No. DMR 1720256; a member of the NSF-funded Materials Research Facilities Network. S.D.W authors acknowledge support from the Nanostructures Cleanroom Facility at the California NanoSystems Institute (CNSI). G.R. also acknowledges partial support through DOE grant no. DE-SC0019166.

**Author contributions**

A.R. and D.H. planned the experiment. A.R., H. N and G.Z. performed the measurements. A.R., G.Z., R.D.A. and D.H. analyzed the data. S.C. and G.R. performed the theoretical calculations. E.Z. and S.D.W. prepared and characterized the samples. A.R., S.C. and D.H. wrote the manuscript.

**Additional information**

Supplementary information is available in the online version of the paper. Correspondence and requests for materials should be addressed to D.H. (dhsieh@caltech.edu).

**Figure 1**

**Ultrafast exchange enhancement and spin–DECP mechanisms. a**, Schematic showing the triplet and singlet spin configurations of our toy model before (bottom row) and after (top row) CT excitation. The superexchange energy $J_{ex}$, defined as the triplet and singlet energy splitting, is a function of the metal *d*-orbital to ligand $p_{x,y}$ orbital hopping energy *t*, the charge transfer energy $\Delta_{CT}$ and the Hund's coupling $J_H$ between the ligand *p*-orbitals. Owing to the bond angle θ being near 90°, hopping from the left (right) metal site to the $p_y$ ($p_x$) ligand orbital is neglected. Expressions for $J_{ex}$ derived from a perturbative calculation (Supplementary Section S1) in both the ground state (GS) and CT state are displayed to the right. **b**, Schematic of the potential energy landscape of the nuclei as function of θ. The CT excitation (dotted red line) causes a sudden shift in the potential minimum due to an impulsive change in $J_{ex}$, launching coherent phonon oscillations about the new minimum.

**Figure 2**

**Spin-DECP detection method in CrSiTe$_3$. a**, Schematic of CrSiTe$_3$ density of states (DOS) showing that excitation with 1 eV light creates a hole in the Te 5*p* valence band and an electron in the Cr $e_g$ conduction band. This excitation has the dual effect of reducing the electrostatic attraction between the Cr and Te ions and also enhancing $J_{ex}$, which acts to decrease and increase the Cr-Te-Cr bond angle θ respectively, in turn launching the $A_g^2$ phonon mode with opposite initial directions as shown in the left panel. A schematic of the pump-probe optical reflectivity experiment used to detect the phase of the coherent $A_g^2$ mode is shown below. **b**, (Top) Transient reflectivity of CrSiTe$_3$ following a 1 eV pump pulse taken at *T* = 25 K. Inset shows the Fourier transform of the data with a clear peak at the $A_g^2$ mode frequency. (Bottom) Oscillatory component of the reflectivity transients ($\Delta R/R_{osc}$) taken at *T* = 25 K (blue) and *T* = 125 K (orange, vertically offset for clarity) obtained by subtracting a fitted exponential background (dashed orange line in top panel) from the raw data and then applying a moving average filter.

**Figure 3**

**CT excitation of $A_g^2$ mode. a**, Temperature dependence of the *xxxz* component of the electric quadrupole SHG response of CrSiTe$_3$, which is known to track the in-plane spin correlator $\langle \vec{S}_i \cdot \vec{S}_j \rangle$[19]. Spin correlations markedly increase below $T_{2D}$ ~ 110 K and then diverge at $T_c$ = 33 K when long-range order sets in. **b**, Temperature dependence of the normalized oscillatory component of the reflectivity transients ($\Delta R/R_{osc}$) taken with 1 eV pump excitation after application of a moving average filter. **c**, Temperature dependence of the $A_g^2$ phonon oscillation phase (top) and amplitude (bottom) extracted from the $\Delta R/R_{osc}$ data. The former was extracted through fitting (Supplementary Section S2) and the red line is a guide to the eye. The latter was extracted from the FFT amplitude, with a positive or negative sign denoting a 0 or π oscillation phase respectively. Error bars on the phase and amplitude data are estimated based on the accuracy of determining peak locations in the $\Delta R/R$ data and the FFT spectra respectively.

## Figure 4

**Sub-gap excitation of $A_g^2$ mode. a**, Transient reflectivity of CrSiTe$_3$ following a 0.14 eV pump pulse taken at $T$ = 25 K (blue) and $T$ = 100 K (orange) showing oscillations at the $A_g^2$ mode frequency. Inset shows the temperature dependence of the $A_g^2$ phonon oscillation phase extracted using the same method as described for the 1 eV pump case. Error bars are estimated based on the accuracy of determining peak locations in the $\Delta R/R$ data. **b**, Oscillatory component of the reflectivity transients shown in panel a obtained by subtracting a fitted exponential background (dashed orange line in panel a) from the raw data. Orange curve is vertically offset for clarity.

## Methods

**Sample growth and characterization**. The CrSiTe$_3$ crystals used in this study were grown using a Te self-flux technique [18]. High purity Cr (Alfa Aesar, 99.999%), Si (Alfa Aesar 99.999%) and Te (Alfa Aesar 99.9999%) were weighed in a molar ratio of 1:2:6 (Cr:Si:Te) and loaded into an alumina crucible sealed inside a quartz tube. The quartz ampoule was evacuated and then backfilled with argon before sealing. Plate-like crystals up to 5 mm thick with flat, highly reflective surfaces were then removed from the reaction crucible. X-ray diffraction (XRD) data collected on crushed crystals and *c*-axis oriented flakes using an Emperyan diffractometer (Panalytical) confirmed the correct $R3$ space group (number 148), CrSiTe$_3$ phase. Measurements of the magnetic susceptibility using a SQUID magnetometer (MPMS, Quantum Design) also confirmed the correct phase as well as sample quality, with $T_c$ ~ 33 K. Samples were cleaved prior to optical measurements and immediately pumped down in an optical cryostat to a pressure better than 10$^{-6}$ Torr.

**Transient reflectivity measurements**. The near-IR pump of 1.2 μm (1 eV) was generated by an optical parametric amplifier (OPA) pumped by a femtosecond titanium-sapphire laser amplifier (repetition rate, 100 kHz; pulse energy, 8 μJ; pulse duration, 100 fs; central wavelength, 800 nm). The pump fluence was 0.25 mJ/cm$^2$. The mid-IR pump of 9 μm (0.14 eV) was generated by the signal outputs from two collinear dual-stage OPAs through the difference frequency generation in a GaSe crystal. The OPAs were pumped by a femtosecond titanium-sapphire laser amplifier (repetition rate, 1 kHz; pulse energy, 6 mJ; pulse duration, 100 fs; central wavelength, 800 nm). The pump fluence was varied between 2 – 8 mJ/cm$^2$ and the transient reflectivity traces were verified to scale linearly with fluence over this range. The dataset shown in Fig. 4 was for 8 mJ/cm$^2$. For both the 1 eV and 0.14 eV pump experiments, the probe pulse was the fundamental output of the amplifiers.

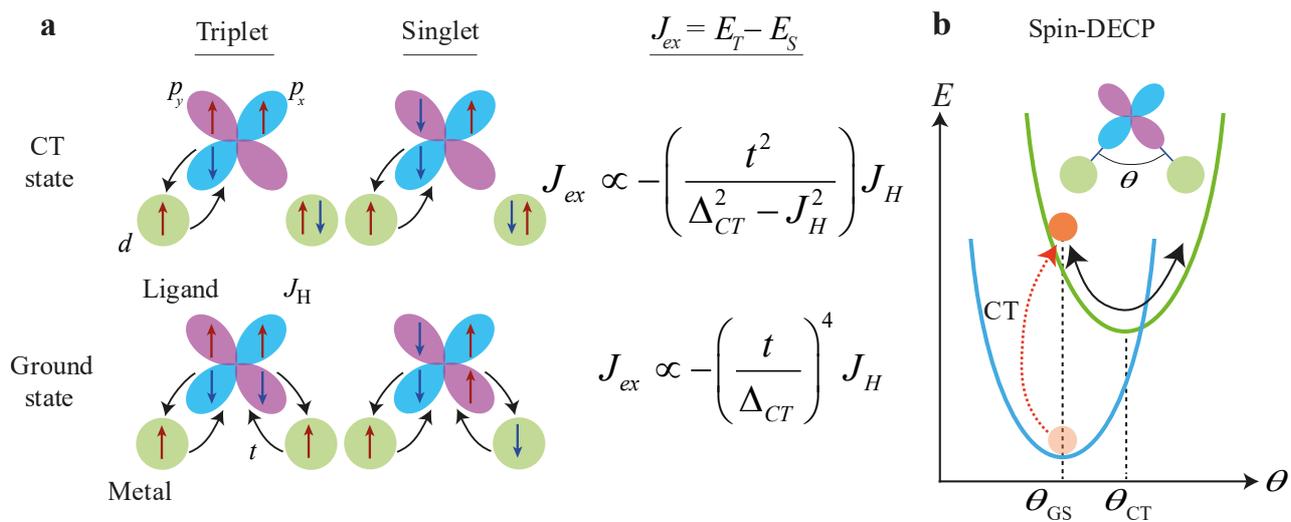

Fig. 1

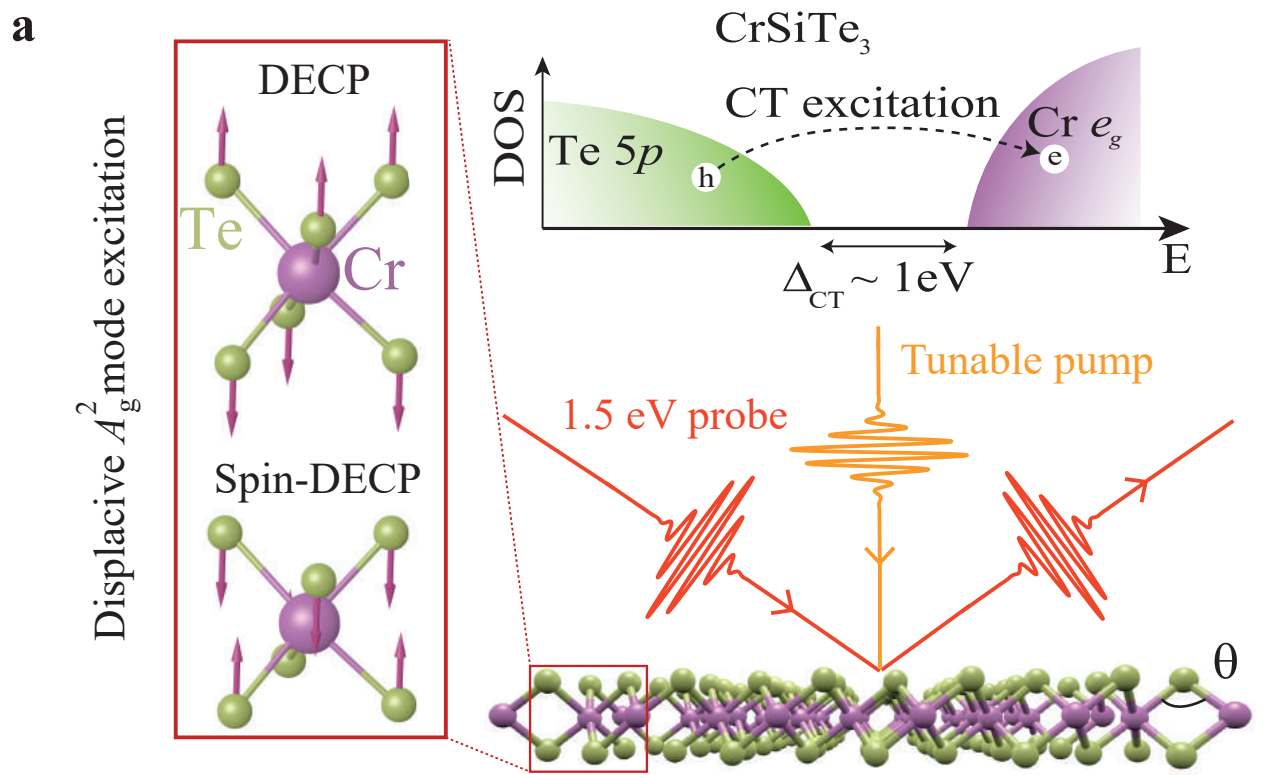

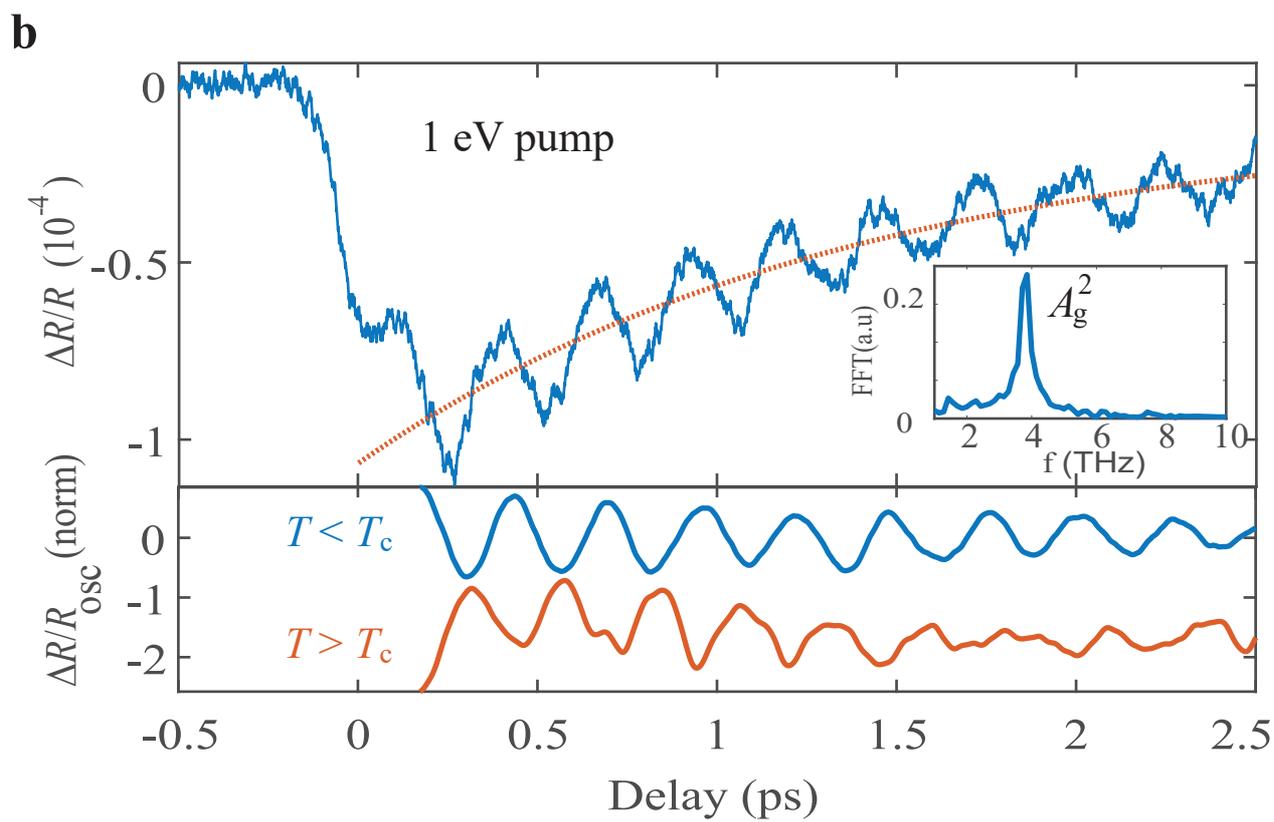

Fig. 2

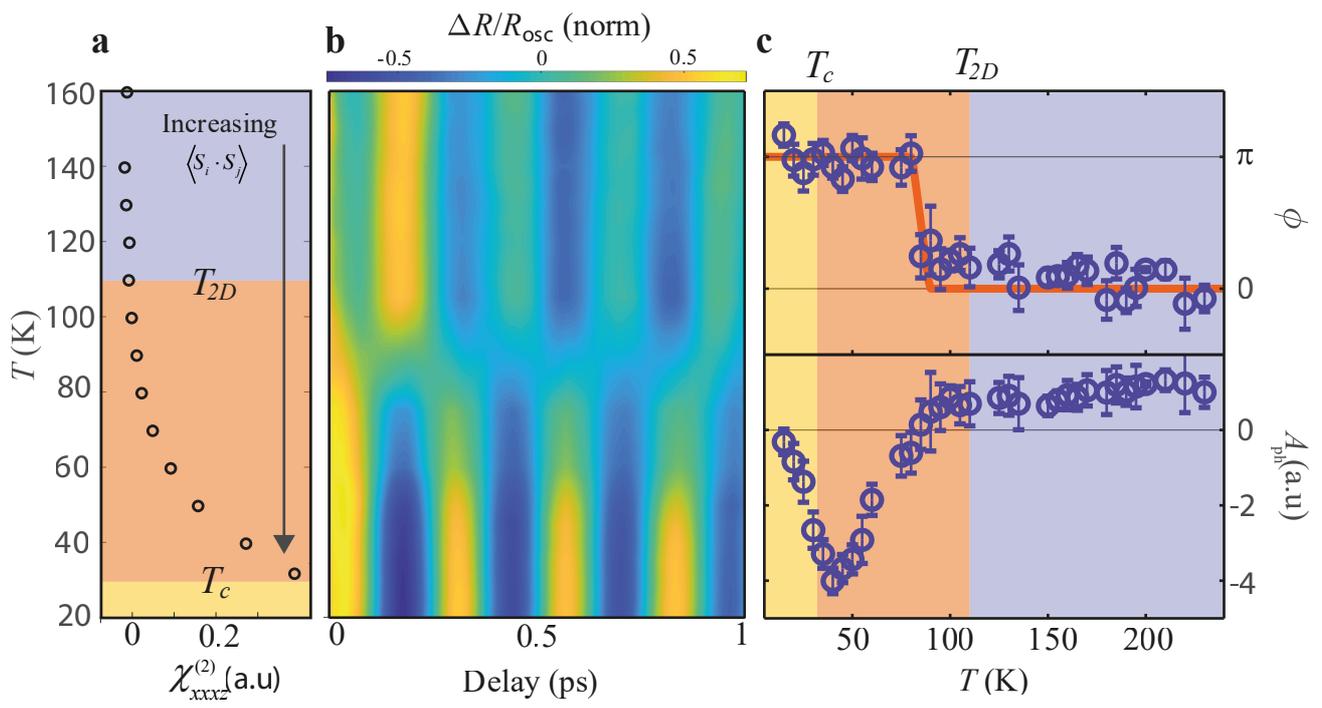

Fig. 3

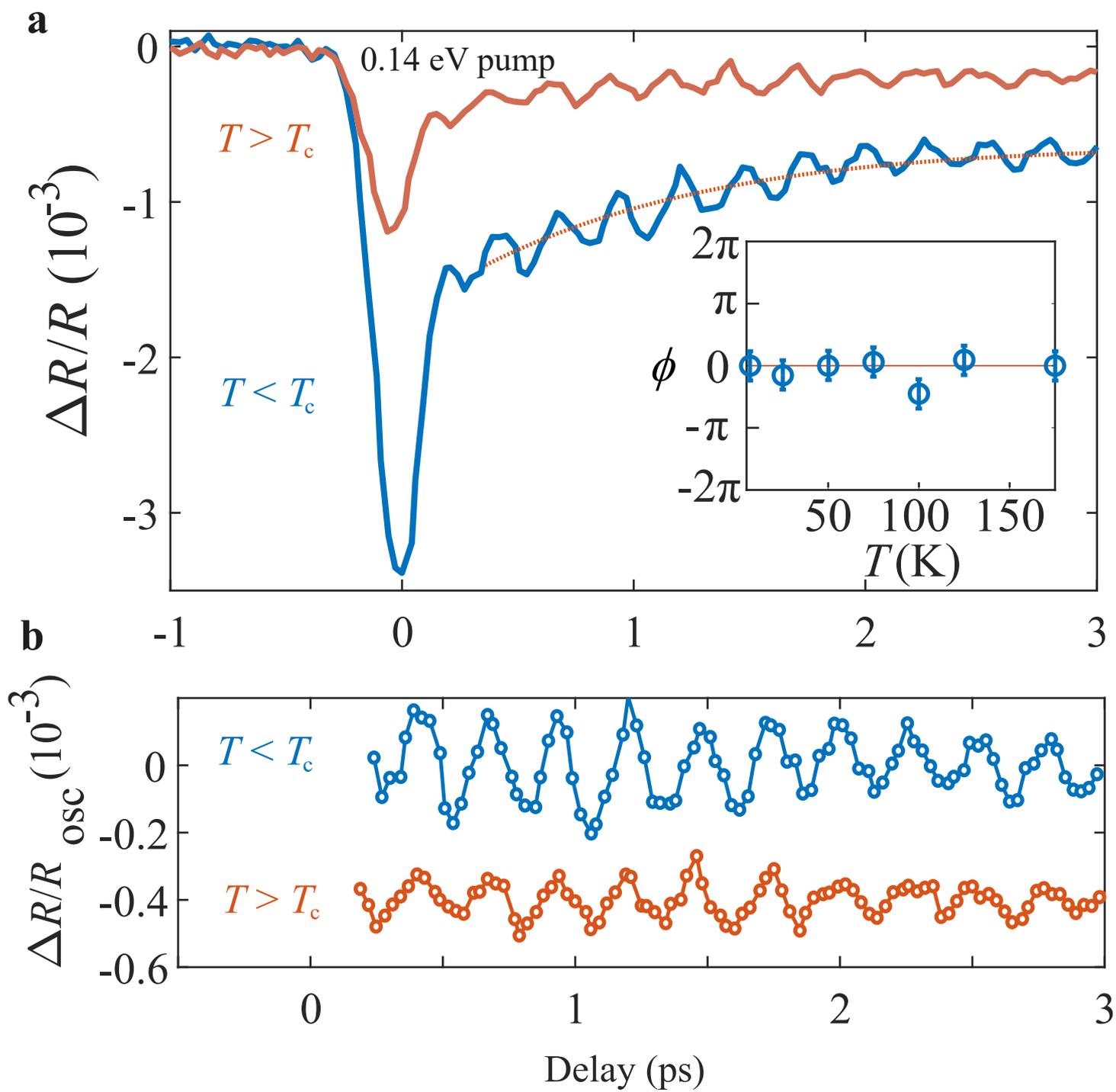

Fig. 4

Supplementary Information for

# Ultrafast enhancement of ferromagnetic spin exchange induced by ligand-to-metal charge transfer

A. Ron et al.

Contents:

# S1. Details of perturbative calculations

## (a) Calculation of $J_{ex}$ before and after CT excitation

We consider a toy molecular model consisting of two metal ions ($i = 1, 2$) and one ligand ion, with a metal-ligand-metal bond angle close to 90°. We represent each metal ion by one singly occupied $d$-orbital and the ligand ion by two doubly occupied $p$-orbitals (X and Y). The $p$-orbitals mediate ferromagnetic superexchange interactions between the $d$-orbitals of the metal ions, which arise due to virtual hopping processes between the metal and ligand orbitals with hopping energies $t_{X1}$, $t_{Y1}$, $t_{X2}$ and $t_{Y2}$ (Fig. S1). Owing to the near 90° bond, we will assume that the X(Y) orbital only has overlap with the $d$-orbital on metal ion 1(2) such that we can set $t_{X1} = t_{Y2} \equiv t$ and $t_{X2} = t_{Y1} = 0$. We also assume a Hund's coupling $J_H$ between the two $p$-orbitals of the ligand ion.

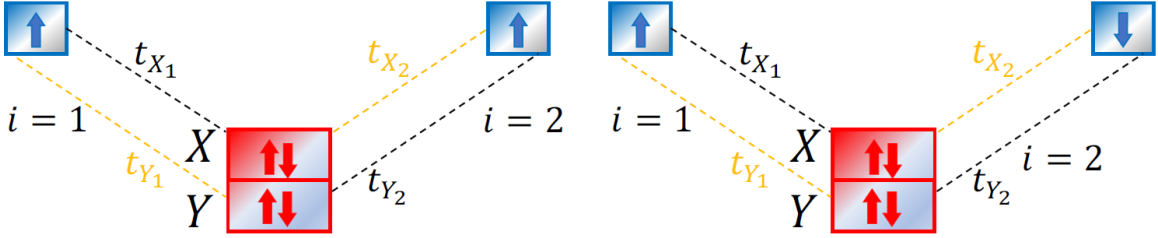

*Figure S1. Toy model consisting of two metal ions (i = 1,2) interacting via superexchange mediated by a ligand ion. The two ligand orbitals are denoted X and Y. The (left) triplet and (right) singlet configurations before CT excitation are shown. Dashed lines show the possible virtual hopping paths, with black (yellow) colors denoting paths with strong (weak) orbital overlap.*

Following the formalism outlined in Chapter 7 of Pavarini *et al*[1], we explicitly calculate the superexchange interaction mediated by the ligand ion within our toy model, defined as the energy difference between the spin triplet and singlet states ($J_{ex} = E_T - E_S$), before and after charge transfer (CT) excitation. We ignore the antiferromagnetic direct exchange between the two metal sites, which is not affected by the metal-ligand interactions.

In the equilibrium state before charge transfer (BCT), Ref.[1] showed that only fourth-order virtual processes contribute to superexchange, which is given by:

$$J_{ex}^{BCT} = E_T - E_S = -\frac{4t^4}{\Delta_{CT}^2} \frac{2J_H}{4\Delta_{CT}^2 - J_H^2}$$

where the charge transfer gap $\Delta_{CT} = U_d + \varepsilon_d - \varepsilon_p$ is defined as the sum of the Coulomb repulsion energy between two electrons in a metal $d$-orbital ($U_d$) and the energy difference between the metal $d$- and ligand $p$-orbitals $(\varepsilon_d - \varepsilon_p)$. For typical values of $\Delta_{CT} \sim 1$ eV and $J_H \sim$ 0.5 eV found in transition metal oxide and chalcogenide materials, we make the approximation that $4\Delta_{CT}^2 - J_H^2 \approx 4\Delta_{CT}^2$, which yields the expression shown in the main text:

$$J_{ex}^{BCT} = -2\left(\frac{t}{\Delta_{CT}}\right)^4 J_H$$

Using appropriate estimates of $\Delta_{CT} \sim 1$ eV, $t \sim 0.5$ eV and $J_H \sim 0.5$ eV for CrSiTe$_3$, Eqn. 2 gives $J_{ex}^{BCT} \approx 60$ meV, which is the same order of magnitude as that reported for CrSiTe$_3$ using density functional theory calculations[2].

The triplet and singlet configurations of the system after charge transfer (ACT) are shown in Fig. S2. In contrast to the BCT case, one can directly see that the energy of a second-order virtual hopping process between metal ion 1 and the ligand ion depends on the initial spin configuration due to Hund's coupling on the ligand site. This produces a finite second-order contribution to the superexchange that is absent in the BCT case.

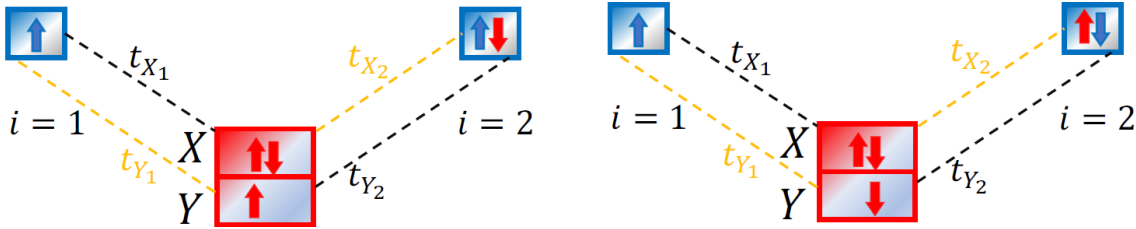

*Figure S2. (left) Triplet and (right) singlet configurations after CT excitation are shown. We ignore the case where an electron is excited from the ligand Y orbital to metal site 1 due to the 90° bond angle.*

More explicitly, the triplet energy is given by:

$$E_T = -\frac{t_{X1}^2}{\Delta_{CT} - J_H} + \frac{t_{Y2}^2}{\Delta_{CT}}$$

while the singlet energy is given by:

$$E_S = -\frac{t_{X1}^2}{\Delta_{CT} + J_H} + \frac{t_{Y2}^2}{\Delta_{CT}}$$

As a result, we arrive at the expression for the triplet-singlet energy splitting that is shown in the main text:

$$J_{ex}^{ACT} = -2\left(\frac{t^2}{\Delta_{CT}^2 - J_H^2}\right)J_H$$

Using again the previous values of $\Delta_{CT} \sim 1$ eV, $t \sim 0.5$ eV and $J_H \sim 0.5$ eV for CrSiTe$_3$ we find that $J_{ex}^{ACT} \approx 300$ meV, a roughly 5-fold increase compared to the BCT case where the spin-dependent triplet-singlet energy difference was only sensitive to fourth-order virtual processes.

**(b) Multi *d*-orbital effects**

The toy model considered above very well captures the change in the superexchange energy between the BCT and ACT states in CrSiTe$_3$. A more detailed model may account for the fact that the Cr$^{3+}$ ion has three unpaired electrons in $t_{2g}$ orbitals, and the fact that CT excitation moves an electron from the Te *p*-orbital to Cr $e_g$ orbital, but the main effects derived from our single *d*-orbital model remain unchanged as we explain below.

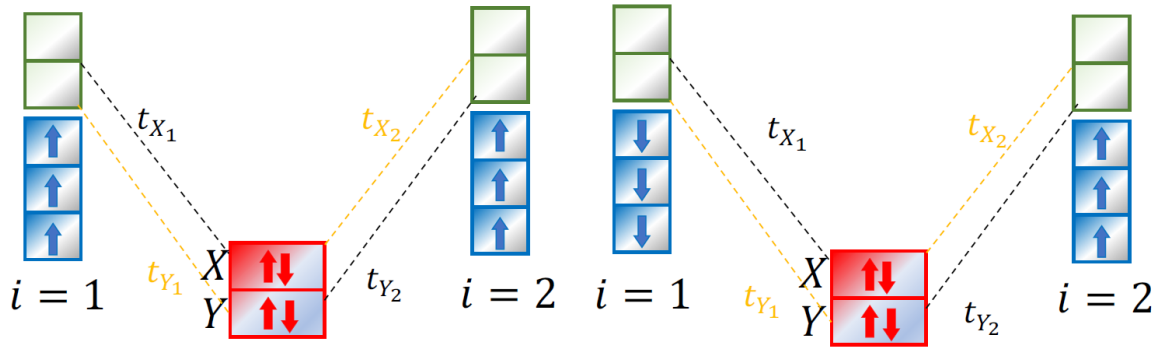

*Figure S3. Multi-orbital model consisting of two metal ions each with three unpaired spins in $t_{2g}$ orbitals (shown in blue) and fully occupied ligand orbitals X and Y (shown in red). This figure shows the (left) triplet and (right) singlet configurations before the charge transfer process. As before, we assume a near 90° bond angle such that the metal 1(2) orbitals have a very large overlap with the ligand X(Y) orbital.*

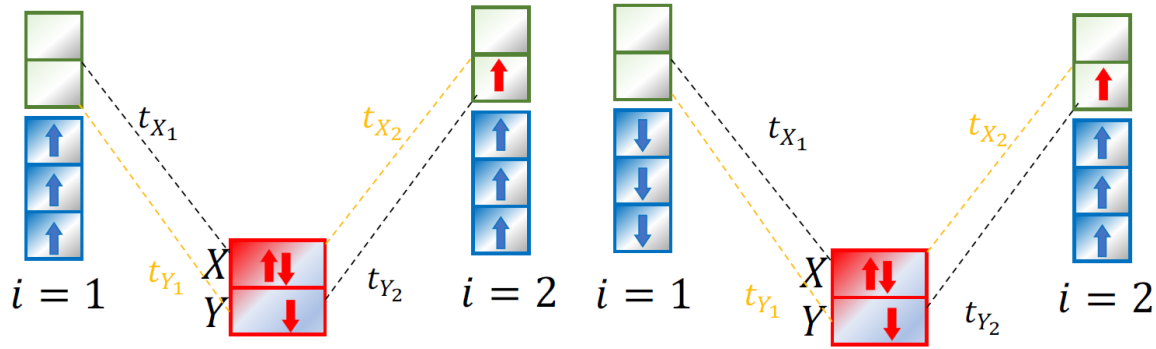

*Figure S4. (left) triplet and (right) singlet cases after the charge transfer process. This is one of the two possibilities for charge transfer associated with 1 eV excitation in CrSiTe$_3$ where an electron is moved from the Y to $e_g$ orbital (shown in green) at site 2. In this case, the transferred spin is parallel to the spin of the unpaired electrons at the metal 2 site because of Hund's coupling within the d-orbitals.*

In the BCT state of the multi-orbital model (Fig. S3), the virtual hopping energies are dictated by Hund's coupling within the ligand *p*-orbitals as well as the strong Hund's coupling within the *d*-orbitals. Therefore the ligand mediated superexchange interactions again favor a FM state with a strength that depends on fourth-order virtual hopping processes. In the ACT state (Fig. S4), the spin of the electron transferred from ligand to metal depends on the spin of the other unpaired electrons present in the $t_{2g}$ orbitals due to the large Hund's coupling within the *d*-orbitals. This is corroborated by first principles calculations[3] that show it is the parallel spin $e_g$ states that lie above the charge transfer gap of ~ 1 eV. The opposite spin states lie much higher in energy due to Hund's coupling within the *d*-orbitals. Therefore the unpaired spin left on the ligand Y orbital after CT excitation must be anti-parallel to the spins on the $t_{2g}$ orbital of metal site 2. Similar to our single *d*-orbital model (Fig. S2), virtual hopping processes between the ligand X orbital and metal site 1 then become sensitive to the initial spin configuration by virtue of Hund's coupling on the ligand site, once again introducing second-order contributions to the superexchange.

**(c) Numerical force estimates for CrSiTe$_3$**

***Exchange force***. The total change in superexchange coupling before and after CT excitation in our single *d*-orbital model is given by:

$$\Delta J_{ex} = J_{ex}^{ACT} - J_{ex}^{BCT} = -2\left(\frac{t^2}{\Delta_{CT}^2 - J_H^2}\right)J_H + 2\left(\frac{t}{\Delta_{CT}}\right)^4 J_H$$

The exchange energy depends on magnetic coupling and spin correlations. This gives rise to an exchange force $\vec{F}_{ex}$ that can be calculated by evaluating the gradient of the change in exchange energy $\Delta E_{ex}$ as:

$$\Delta E_{ex} = \Delta J_{ex} \langle \vec{S}_1 \cdot \vec{S}_2 \rangle$$

and

$$\vec{F}_{ex} = \vec{\nabla}(\Delta E_{ex})$$

For CrSiTe$_3$, we are specifically interested in understanding the force along the $A_g^2$ phonon coordinate *y*, the direction along the perpendicular from the Te ion to the line joining two Cr ions. This depends on how the hopping parameters change as a function of *y*, with *y* = 0 being the equilibrium value. To quantitatively estimate $\vec{F}_{ex}$, we compute the *y* dependence of the hopping parameters using Slater type orbitals[4] – with an effective charge slightly higher than the one given by the Slater method in order to match the expected value of hopping parameter – and the reported bond lengths for CrSiTe$_3$. In agreement with the intuitive arguments outlined in the main text, our calculations (Fig. S5) show that $\vec{F}_{ex}$ tries to increase the bond angle from its equilibrium value < 90° and thus pushes the Te ion towards the Cr ions. To calculate the magnitude of the force, we keep only the second-order contribution to $\Delta J_{ex}$ and use a quantitative estimate for the gradient of the hopping energy based on our Slater orbital model ($\partial_y t^2|_{y=0} \approx -1$

eV$^2$/Å) (Fig. S5) together with the previously used values of $\Delta_{CT} \sim 1$ eV, $t \sim 0.5$ eV and $J_H \sim 0.5$ eV for CrSiTe$_3$. Plugging this into our expression for $\vec{F}_{ex}$ yields:

$$F_{ex}^y \approx -\frac{1}{3}\langle \vec{S}_1 \cdot \vec{S}_2 \rangle$$

in units of eV/Å.

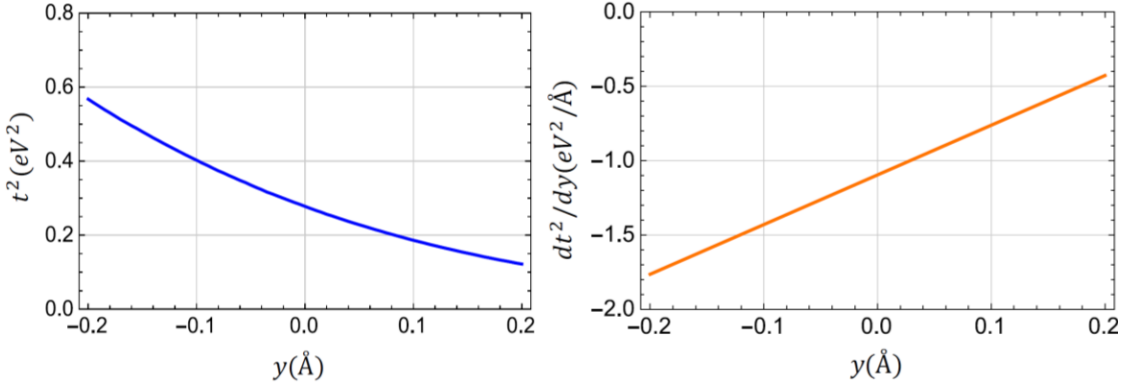

*Figure S5. Change in hopping parameter t and its derivative as a function of the displacement along y direction calculated using a Slater orbital model.*

***Coulomb force***. Next we estimate the additional Coulombic force $\vec{F}_c$ on a Te atom in the ACT state compared to the BCT state due to changes in the spatial charge distribution. Based on the reported local magnetic moment value of 3.87 $\mu_B$ on the Cr$^{3+}$ ions[3] in the BCT state, we estimate that there is an extra –0.87$e$ charge on the Cr ions, which is reportedly due to Cr-Te *dp*-hybridization. Therefore we infer that the charge on the Cr ion is +2.13$e$. The extra charge of +0.87$e$ is shared among three Te$^{2-}$ ions based on the CrSiTe$_3$ stoichiometry, and thus each Te atom has an extra charge of +(0.87/3)$e$, making the charge on each Te ion equal to (–2 + 0.87/3)$e$ ~ –1.7$e$. For simplicity, we assume that the CT process involves the transfer of one electron from a Te ion to its neighboring Cr ion, which are located by the small black circles in Fig. S6. For the purpose of this simple calculation we ignore the changes in charge distribution on other Cr and Te ions inside the larger circle in Fig. S6. The vertical (*y*) component (perpendicular to the honeycomb plane) of the Coulombic force $F_c^y$ on the encircled Te ion arises mainly due to a reduction in attractive forces from oppositely charged Cr ions. However, a modified charge on the encircled Te ion also affects the Coulombic repulsion from other Te ions. In order to estimate $\vec{F}_c$ after the CT process, we treat all these ions as point charges and calculate $\vec{F}_c$ on the encircled Te ion using the expression:

$$\vec{F}_c = \sum_i \frac{q_{Te}^f q_i^f - q_{Te}^0 q_i^0}{r_i^2} \hat{r}_i$$

where $q_i^f$ and $q_i^0$ denotes the charge on the $i^{th}$ ion in the ACT and BCT state respectively, $\vec{r}_i$ is a vector pointing from the $i^{th}$ ion to the Te ion, and the index *i* runs over all the Cr and Te atoms inside the larger circle in Fig. S6. In this approximation, we calculate $F_c^y \sim +2$ eV/Å, denoting repulsion of the encircled Te ion away from the Cr layer.

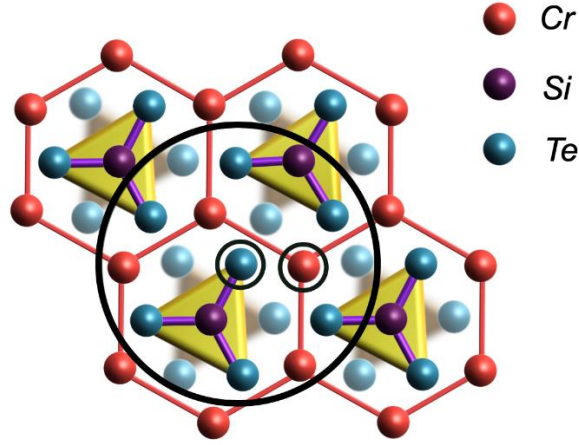

*Figure S6. The top view of a CrSiTe₃ monolayer. Cr atoms are arranged on a honeycomb lattice with a layer of Te atoms above and below. Based on x-ray structure refinements[5], in our model we use a Cr-Te bond length of 2.753 Å where the vertical distance is 1.67 Å and the horizontal distance is 2.189 Å, and a Si-Te bond length of 2.477 Å with vertical distance of 0.567 Å and a horizontal distance 2.4119 Å. The y-direction points out of the page.*

*Net force*. Combining our results for the exchange and Coulomb contributions to the *y*-component of force on the encircled Te ion in Fig. S6 we get the expression:

$$F^y = F_c^y + F_{ex}^y \approx \left(2 - \frac{1}{3}\langle \vec{S}_1 \cdot \vec{S}_2 \rangle\right)$$

We see that the exchange force competes with the Coulombic force and thus the total force should change from being positive to zero to negative as $\langle \vec{S}_1 \cdot \vec{S}_2 \rangle$ increases. This explains the observed trend in the amplitude of the coherent $A_g^2$ phonon mode as spin correlations increase upon cooling. We note that our calculations present a very rough estimate and that the Coulombic force is likely overestimated given that we ignored the shielding effect from electron clouds and also assumed that the CT process involves the transfer of one electron from the Te ion to the Cr ion. Model Hamiltonian calculations have shown that $\langle \vec{S}_1 \cdot \vec{S}_2 \rangle \approx 0.1$ at $T = 90$ K in CrSiTe₃[6], where we observe the coherent $A_g^2$ amplitude and thus $F^y$ to cross zero. This is consistent with the value of $F_c^y$ being overestimated in our model.

## S2. Extracting the phonon phase

The π phase shift in the $A_g^2$ phonon oscillations across $T \sim 90$ K for the 1 eV pump case can directly be seen in the raw data of Fig. S7 or the background subtracted data in Fig. 3b. But we can also quantitatively extract this information as plotted in Fig. 3c. To do so we first extract the frequency ($f$) of the oscillations by plotting the delay time at which each extrema in the $\Delta R/R$ trace appears (Fig. S7). The slope ($a$) of this curve was then obtained by fitting to a straight line, from which $f$ was calculated as $f = 1/(2a)$ since both minima and maxima were plotted. This process was used to obtain $f$ at all temperatures. Once $f$ was obtained the phase of the oscillations was extracted by taking the delay time at which the $n^{th}$ peak (or dip) in the oscillations occurs and multiplying by $2\pi f$.

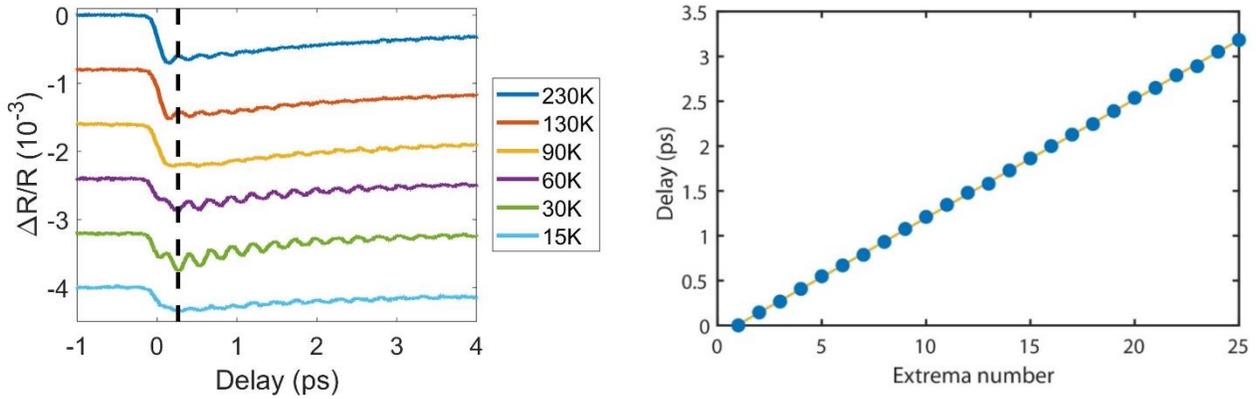

*Figure S7. (Left) Selected raw transient reflectivity traces taken with 1 eV pump photon energy and 1.5 eV probe photon energy. Curves are vertically offset for clarity. Upon cooling, the suppression of the coherent phonon oscillation amplitude around T = 90 K and its revival with opposite phase at lower temperatures can clearly be seen in the raw data. Dashed vertical line is a guide to the eye to highlight the π phase flip. (Right) Time delay values at which the oscillation maxima and minima occur in the 15 K transient reflectivity dataset plotted in order of their appearance. The yellow line is a linear fit from which the frequency was calculated.*

## S3. Ultrafast SHG and polarization rotation based control experiments

To further verify that the $\pi$ phase change in the coherent $A_g^2$ phonon oscillations observed across $T \sim 90$ K for the 1 eV pump case originates from a competition between conventional and spin-DECP effects, we also measured the temperature dependence of a coherent $E_g$ phonon mode upon 1 eV pump excitation. Unlike totally symmetric $A_g$ phonons, the symmetry lowering $E_g$ phonon is not excited via a DECP mechanism but rather via an impulsive stimulated Raman scattering (ISRS) mechanism. Since coherent $E_g$ phonon oscillations were not detectable by transient reflectivity measurements, we performed time-resolved second harmonic generation (SHG) experiments that are known to be sensitive to symmetry lowering modes. As shown in Fig. S8, the raw time-resolved SHG transients show clear oscillations with a frequency of around 2.85 THz, closely matching the reported and calculated $E_g$ phonon frequency[7]. Unlike the $A_g^2$ mode reported in the main text, we observe no obvious change in the phase of the $E_g$ mode from 300 K down to 7 K. The absence of a $\pi$ phase flip in the ISRS excited $E_g$ mode is consistent with our theory that CT induced superexchange enhancement only generates coherent phonons through a DECP mechanism. We note that the $\pi$ phase flip we observed in the transient reflectivity measurements across $T \sim 90$ K cannot be due to a change in excitation mechanism from DECP to ISRS because such a change would yield a $\pi/2$ phase flip and because it is usually caused by changes in sample opacity or dielectric properties[8], which does not apply to CrSiTe$_3$.

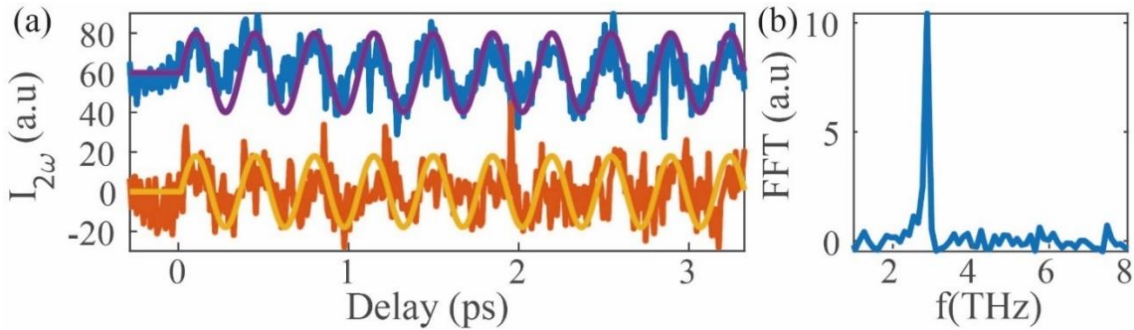

*Figure S8. (a) Transient SHG intensity versus time delay at 300 K (orange) and 7 K (below) following 1 eV pump excitation showing the absence of a phase flip. Solid lines are guides to the eye. Curves are vertically offset for clarity. (b) Fourier transform of a typical low temperature SHG transient showing the frequency of the $E_g$ mode oscillations at $f \sim 2.85$ THz.*

To show that the different temperature dependent behavior of the $A_g^2$ and $E_g$ phonon modes observed via transient reflectivity and SHG is intrinsic and not due to the different probing methods, we carried out time-resolved optical polarization rotation measurements that are able to simultaneously monitor both the $A_g^2$ (DECP) and $E_g$ (ISRS) responses following 1 eV pump. As shown in the inset of Fig. S9a, both the frequencies of the $A_g^2$ and $E_g$ modes are present in the transient polarization rotation data. The temperature dependence of their amplitudes extracted from FFT analysis (Fig. S9b) corroborates our previous findings that the $E_g$ mode amplitude is temperature independent while the $A_g^2$ mode amplitude crosses zero at $T \sim 90$ K. We note that the pump fluence used in the transient optical rotation measurements was an order of magnitude larger compared to that used for the transient reflectivity measurements shown in the main text. The fact that the $A_g^2$ mode goes to zero at the same temperature in both cases (Fig. S9) is further evidence that the magnitude of the superexchange enhancement does not depend on the pump electric field, supporting the CT induced enhancement picture and ruling out a Floquet engineering mechanism (see Section S5).

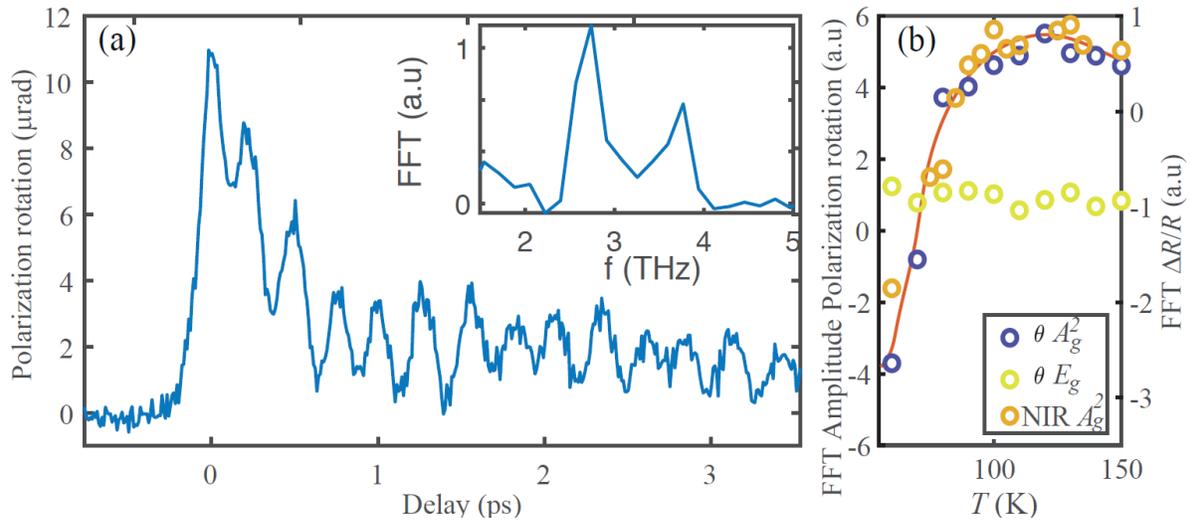

*Figure S9. (a) Transient polarization rotation ($\theta$) trace taken with 1 eV pump photon energy at T = 15 K. The inset is a Fourier transform showing the two peaks at the $A_g^2$ and $E_g$ phonon mode frequencies. (b) The temperature dependence of the $A_g^2$ and $E_g$ mode amplitudes extracted from the transient polarization rotation measurements (left axis) plotted together with that of the $A_g^2$ mode amplitude extracted from the transient reflectivity measurement shown in Fig. 3 of the main text (right axis).*

## S4. Extended transient reflectivity data with 0.14 eV pump

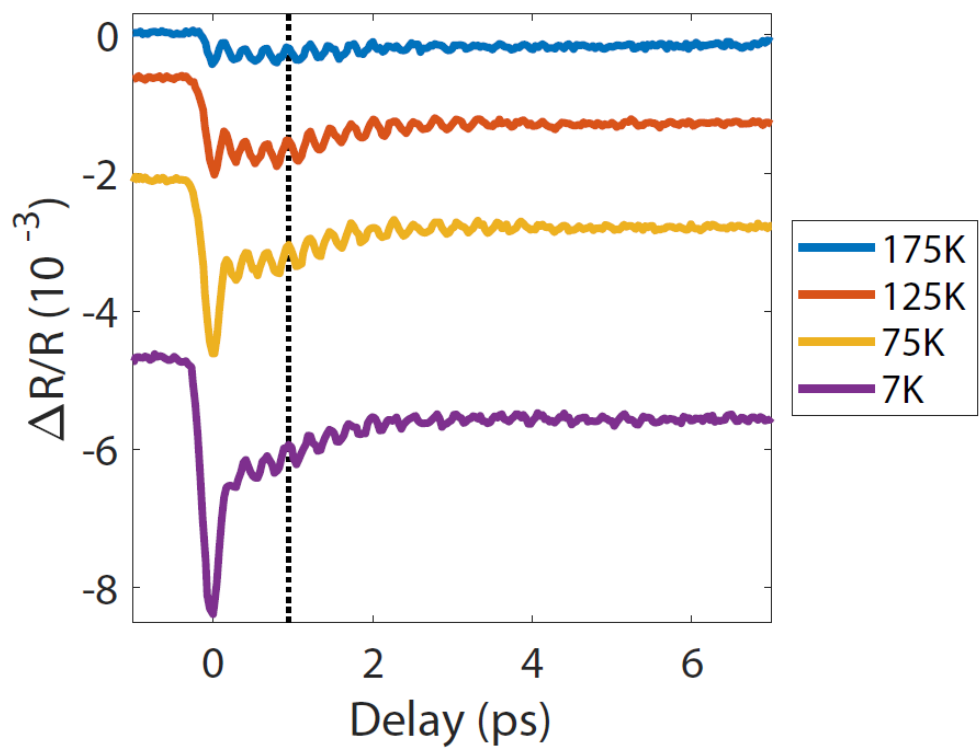

*Figure S10. Raw reflectivity transients at select temperatures taken with 0.14 eV pump photon energy and 1.5 eV probe photon energy. Curves are vertically offset for clarity. Coherent oscillations of the same $A_g^2$ mode as observed in the 1 eV pump data are clearly resolved. Dashed vertical line is a guide to the eye to highlight the absence of any temperature dependent phase shift.*

## S5. Calculation of Floquet engineering effects

A time-periodic electric ($E$) field can modify exchange interactions due to hopping renormalization. In CrSiTe$_3$, the FM interactions arise due to virtual hopping between ligand Te ions and metal Cr ions. The two magnetic site model introduced in Section S1 can be described by the following tight-binding Hamiltonian:

$$H = U_d \sum_{i=1,2} \hat{n}_{i\uparrow}\hat{n}_{i\downarrow} + J_H \sum_{\substack{\alpha=X,Y \\ \alpha \neq \alpha'}} c^\dagger_{\alpha\uparrow} c^\dagger_{\alpha'\downarrow} c_{\alpha\downarrow} c_{\alpha'\uparrow} + \sum_{\substack{\alpha=X,Y \\ \sigma=\uparrow,\downarrow}} \varepsilon_p \hat{n}_{\alpha\sigma} - \sum_{\substack{i=1,2 \\ \alpha=X,Y}} t_{\alpha i}(c^\dagger_{i\sigma} c_{\alpha\sigma} + c^\dagger_{\alpha\sigma} c_{i\sigma})$$

where $i = 1, 2$ are the two Cr sites, $X, Y$ denote the two orbitals of the Te ion, $\varepsilon_p$ is the electronic energy of the Te $p$-orbitals, $J_H$ is the Hund's coupling between these orbitals, and $U_d$ denotes the Coulombic repulsion for electrons in the Cr $d$-orbitals. In this case, the charge transfer gap $\Delta_{CT} = U_d - \varepsilon_p$, where all energies are measured with respect to the energy of Cr $d$-orbitals. As before, owing to the 90° bond angle, we $t_{X1} = t_{Y2} \equiv t$ and $t_{X2} = t_{Y1} = 0$. The effect of a circularly polarized A.C electric field can be incorporated by making a Peierls substitution that results in photo-assisted hopping, and thus allows virtual excitations to many different Floquet sectors. By applying fourth-order perturbation theory, we calculate the magnetic coupling strength for the Floquet Hamiltonian:

$$J = t^4 \sum_{m_1,n_1,n} \frac{4J_H}{(\Delta_{CT} + m_1\omega)(\Delta_{CT} + n_1\omega)((2\Delta_{CT} + n\omega)^2 - J_H^2)} \times$$

$$\big(\cos(2(m_1 - n_1)\alpha)\mathcal{J}_{n_1}(\zeta_{X1})\mathcal{J}_{m_1}(\zeta_{X1})\mathcal{J}_{n-n_1}(\zeta_{Y2})\mathcal{J}_{n-m_1}(\zeta_{Y2}) \\ + \cos(2(n - m_1 - n_1)\alpha)\mathcal{J}_{n_1}(\zeta_{X1})\mathcal{J}_{m_1}(\zeta_{Y2})\mathcal{J}_{n-n_1}(\zeta_{Y2})\mathcal{J}_{n-m_1}(\zeta_{X1})\big)$$

where $\mathcal{J}_n$ is the $n^{\text{th}}$ order Bessel function and the drive parameter is $\zeta_{X1} = -\zeta_{Y2} = \frac{\zeta_0}{2\cos\alpha}$, with $\zeta_0 = eEa/\hbar\omega$, where $e$ is the electron charge, $a$ is the separation between two Cr ions, $\hbar\omega$ is the driving photon energy, and $\alpha$ is the angle between the line joining the two Cr ions and the component of the Cr-Te bond in the plane containing the Cr ions. The resulting changes in FM coupling are shown in Fig. S11, which show that such Floquet engineering effects are significant only for $E \sim 1$ V/Å. In our 0.14 eV pump experiments, the pump fluence is of order 10 mJ/cm$^2$ and the pulse width is roughly 100 fs, giving a field strength $E \sim 0.1$ V/Å, which corresponds to a negligible enhancement in the magnetic coupling strength ($< 1$ %).

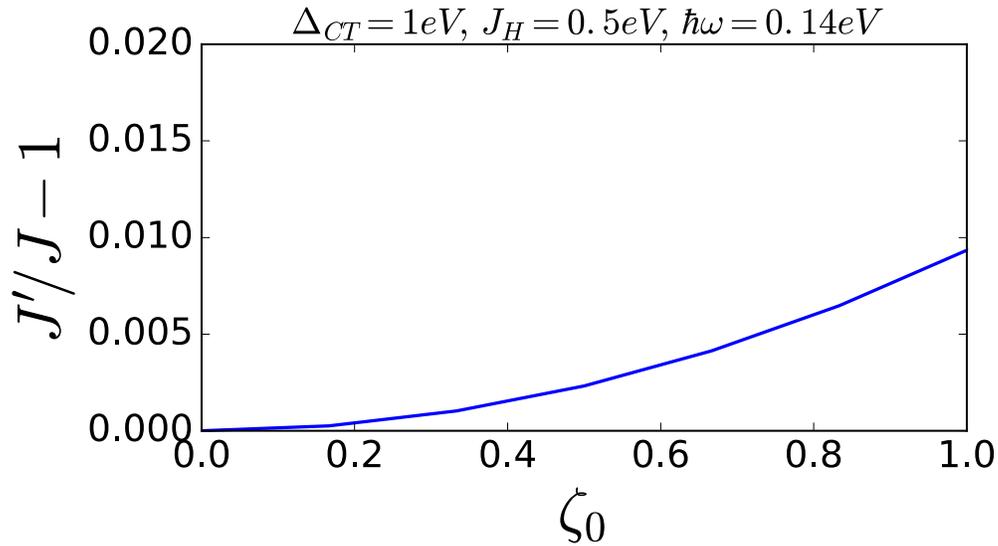

*Figure S11. Calculated fractional change in FM exchange energy without (J) and with driving (J') as a function of the Floquet parameter $\zeta_0$. Summation was carried out over the range $-12 \leq n, n_1, m_1 \leq +12$. An electric field of $E \approx 0.1$ V/Å corresponds roughly to $\zeta_0 = 1$, and thus imparts a change of only about 1 % to the exchange.*

## Supplementary References